\documentstyle[multicol,aps,prl,twocolumn,epsf]{revtex}
\newcommand{\bm}[1]{\mbox{\boldmath $#1$}}
 
\begin{document}
\twocolumn[\hsize\textwidth\columnwidth\hsize\csname@twocolumnfalse\endcsname
\title{Universality and saturation of intermittency
in passive scalar turbulence}
\author{A.~Celani$^{1,2}$, A.~Lanotte$^{1}$, A.~Mazzino$^{3,4}$ 
and M.~Vergassola$^{1}$ \\
\small{$^{1}$ CNRS, Observatoire de la C\^ote d'Azur, B.P. 4229,
06304 Nice Cedex 4, France.}\\
\small{$^{2}$ Dipartimento di Fisica Generale, Universit\`a di Torino,
and INFM Unit\`a di Torino Universit\`a, 
I--10126  Torino, Italy.}\\
\small{$^{3}$ INFM--Dipartimento di Fisica, Universit\`a di Genova,
Via Dodecaneso, 33, I-16142 Genova, Italy.}\\
\small{$^{4}$ The Niels Bohr Institute, Blegsdamvej, 17, 
Copenhagen, Denmark.}}
\date{\today}
\maketitle
\begin{abstract}
The statistical properties of a scalar field advected by the
non-intermittent Navier-Stokes flow arising from a two-dimensional
inverse energy cascade are investigated. The universality properties
of the scalar field are directly probed by comparing the results
obtained with two different types of injection mechanisms. Scaling
properties are shown to be universal, even though anisotropies
injected at large scales persist down to the smallest scales and local
isotropy is not fully restored. Scalar statistics is strongly
intermittent and scaling exponents saturate to a constant for
sufficiently high orders. This is observed also for the advection by a
velocity field rapidly changing in time, pointing to the genericity of
the phenomenon.  The persistence of anisotropies and the saturation
are both statistical signatures of the ramp-and-cliff structures
observed in the scalar field.
\end{abstract}
]

Ramp-and-cliff structures are a characteristic feature
of fields, like dye concentration or temperature, obeying the passive
scalar equation (see, e.g., Refs.~\cite{SS99,UF95})\,:
\begin{equation}
\partial_t T({\bm r},t)+{\bm v}({\bm r},t)\cdot {\bm \nabla}
T({\bm r},t)=\kappa\Delta T({\bm r},t),
\label{scalar}
\end{equation}
i.e. advected by the velocity ${\bm v}$ and smeared out by the
molecular diffusivity $\kappa$. Scalar gradients tend indeed to
concentrate in sharp fronts separated by large regions of weak
gradients (see Fig.~1). The experimental evidence for ramps and cliffs
is long-standing and massive \cite{CG77,PGM82,KRS91,MW98}.
Furthermore, numerical simulations indicate that scalar structures are
not mere footprints of those in ${\bm v}$ and appear also for
synthetic flow \cite{HS94,AP94}. The presence of ramp-and-cliff
structures raises some important issues about scalar turbulence and
its intermittency properties.  Following Kolmogorov's 1941 theory, it
is indeed usually assumed that turbulence restores universality,
i.e. independence of the large-scale injection mechanisms, and
isotropy at small scales (see Ref.~\cite{UF95}).  The evidence for
scalar turbulence is however that anisotropies find their way down to
the small scales, manifesting in the scalar gradient skewness of
$O(1)$, independently of the P\'eclet number
\cite{CG77,PGM82,KRS91,MW98,HS94,AP94}. This is due to the
preferential alignment of ramp-and-cliff structures with large-scale
scalar gradients, present in most experimental situations. For
structure functions $S_n({\bm r})=\langle \left(T({\bm r})-T({\bm
0})\right)^n\rangle$, this persistence is revealed by normalized odd
orders $S_{2n+1}/S_2^{n+1/2}$ decaying more slowly than the expected
$r^{2/3}$ (see Ref.~\cite{SS99}). Is this experimentally observed
behavior signalling that small scales are fully imprinted by the large
scales and that the universality framework should be discarded
altogether\,?  This is the first issue, raised in
Refs.~\cite{SS99,KRS91}, that we shall investigate in this Letter.
The second is about the consequences of cliffs for high-order
intermittency. Their strength candidates them for the dominant
contributions to strong event statistics and the issue raised in
Ref.~\cite{RHK94} is whether structure function scaling exponents are
then saturating to a constant for high orders $n$.

Numerical simulations are an ideal tool to analyze the previous
questions, allowing to probe universality, by comparing the results
obtained with two different types of injection, and saturation, by
gathering enough statistics to capture strong events. Here, we shall
take for ${\bm v}$ a 2D flow generated by a Navier-Stokes inverse
energy cascade \cite{RHK67}. Universality is then understood as
dependence of scalar properties on the injection mechanisms for this
fixed ${\bm v}$ statistics. The scalar is injected at large scales,
comparable to those where the inverse cascade is stopped by friction
effects, and its properties are investigated in
\narrowtext \vskip -2pt\begin{figure} \epsfxsize=8.5truecm \epsfbox{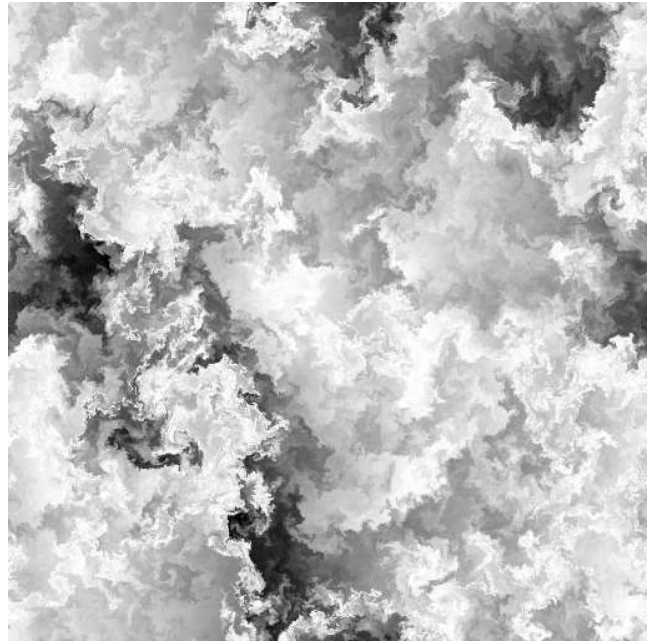}
\vskip 5pt
\caption{A snapshot of the scalar fluctuation field for the injection
by a mean gradient.  Colours are coded according to the deviation from
the average value (from white to black).}
\label{f1} \end{figure}
\narrowtext \begin{figure} \epsfxsize=9truecm \epsfbox{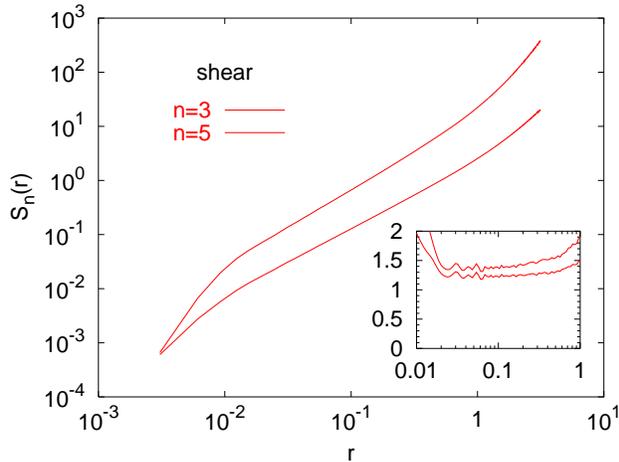}
\caption{The 3rd and the 5th-order parallel structure functions for
the injection by a mean gradient. In the inset, local scaling
exponents $dS_n(r)/d\log r$. The measured exponents are
$\zeta_3=1.25\pm 0.04$, $\zeta_5=1.38\pm 0.07$, with error bars
estimated from rms fluctuations of local scaling exponents.}
\label{f2} \end{figure}
\noindent the energy inertial range (see Ref.~\cite{BCV99} for
details).  There, the velocity is isotropic, scale-invariant with
exponent $1/3$ (no intermittency corrections to Kolmogorov scaling
\cite{PT98,BCV99}) and has dynamical correlation times (finite and
free of synthetic flow pathologies discussed in Ref.~\cite{HS94}).

As for scalar injection, a first choice is naturally suggested by
experiments, where it usually takes place via a large-scale gradient.
We assume then, as in Refs.~\cite{HS94,AP94}, that the average
$\langle T \rangle={\bm g}\cdot{\bm r}$ and we integrate the equation
for the fluctuations $\theta=T-{\bm g}\cdot{\bm r}$,
i.e. (\ref{scalar}) with a source term $-{\bm v}\cdot{\bm g}$ on the
right hand side. A snapshot of the $\theta$ field is shown in
Fig.~1. The presence of the gradient ${\bm g}$ breaks isotropy and
allows for asymmetries and non-vanishing odd-order moments in the
scalar statistics. The second choice is a more artificial random
forcing $f({\bm r},t)$ added to (\ref{scalar}). Its motivation is to
produce an isotropic statistics, e.g. by taking $f$ Gaussian,
with zero average and correlation function $\langle f({\bm
r},t)\,f({\bm 0},0)\rangle=\delta(t)\chi(r/L)$. The scale $L$ where
the injection is concentrated is taken comparable to the velocity
integral scale. The equations for the scalar are integrated in
parallel to the 2D Navier-Stokes equation for about 100 eddy turn-over
times by a standard pseudo-spectral code on a $2048^2$ grid. In the
runs presented in the following, the diffusive term is replaced by a
bi-Laplacian, but it was checked by another series of simulations that
using a Laplacian gives consistent results, although on less extended
scaling ranges.

Let us first show that the persistence of anisotropies observed in
experiments occurs also in our case.  Odd-order structure functions
vanish in the randomly forced case. In the shear case they
do not, except for separations ${\bm r}\perp {\bm g}$. For
non-orthogonal ${\bm r}$'s, the scaling exponents do not depend on the
direction ${\bm r}$ and in Fig.~2 we present the parallel structure
functions, i.e. ${\bm r}$ aligned with ${\bm g}$. The resulting
third-order skewness $S_3/S_2^{3/2}$ scales as $r^{0.25}$, the
\narrowtext \begin{figure} \epsfxsize=9truecm \epsfbox{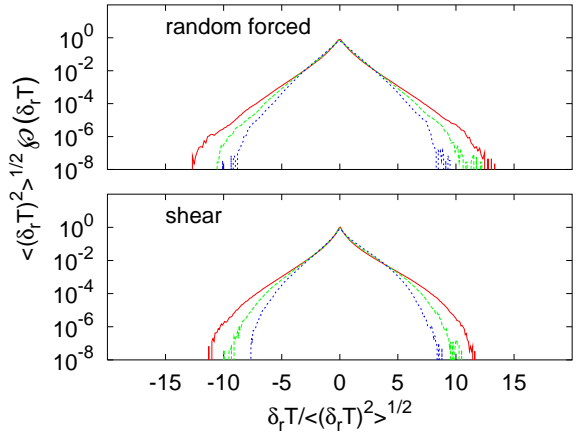}
\vskip 0.2cm
\caption{Pdf's of scalar increments normalized by their standard
deviations for three separations $r=2.5\times\,10^{-2}$,
$5\times\,10^{-2}$, $10^{-1}$ in the inertial range. }
\label{f3}
\end{figure} 

\noindent 2nd-order exponent being $\simeq 2/3$ (see Fig.~4). As in
the experiments, the skewness decay is slower than the expected
$r^{2/3}$. Furthermore, here enough statistics is accumulated to give
access also to the 5th-order. The persistence effect is now dramatic
as $S_5/S_2^{5/2}\sim r^{-0.2}$ increases at small
scales. Intermittency generates of course an ambiguity in the
normalization, e.g.  $S_5/S_4^{5/4}$ is decaying, albeit very
slowly. This reflects the fact that scalar increment pdf's change
shape with $r$ and one should then be specific about which part of it
is sampled and the choice of the observable representative of the
anisotropy degree. It is however unambiguously clear that local
isotropy is not fully restored at small scales and the quality of the
scaling laws found here indicates that this is a genuine effect, not
related to finite P\'eclet numbers.

More insight into this breaking of full universality is gained by
analyzing scalar increment pdf's and moments of even order, which are
non-vanishing for both types of
\narrowtext \begin{figure} \epsfxsize=9truecm \epsfbox{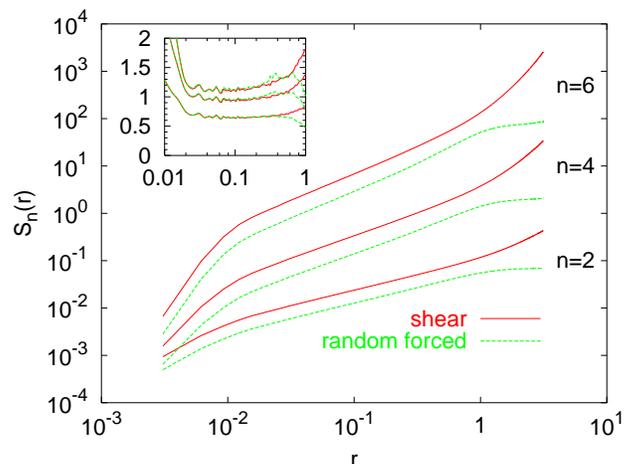}
\caption{Low-order even structure functions. Local scaling exponents
are shown in the inset. The measured exponents are
$\zeta_2=0.66\pm 0.03$, $\zeta_4=0.95\pm 0.04$ and $\zeta_6=1.11 \pm
0.04$.}
\label{f4} \end{figure}
\narrowtext \begin{figure} \epsfxsize=9truecm \epsfbox{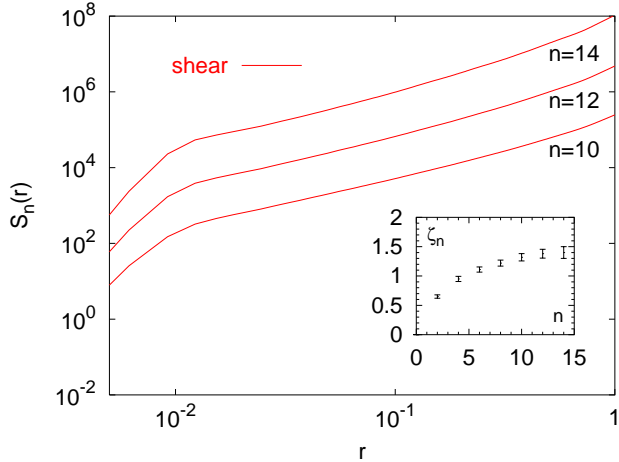}
\caption{Scalar structure functions of orders $10, 12$ and $14$. The
scaling exponents are shown in the
inset.}  \label{f5} \end{figure}
\noindent forcing. Fig.~3 shows that the pdf's for the two types
of injections do not have the same shape (the same holding when
symmetric parts are taken). In the shear case, the separations ${\bm
r}$ have been taken along the diagonal directions, at angles
$\phi=\pi/4$ with respect to ${\bm g}$. This choice is motivated by
the application of the procedure developed in Refs.~\cite{SP,BP} to
the 2D case and permits removal of the first subleading anisotropic
contribution $\propto \cos 2\phi$ to even-order moments. The fact that
the pdf's have different shapes implies that the adimensionalized
constants $C_{n}$ in structure functions $S_n(r)=C_n\left(\epsilon
r\right)^{n/3}\left(L/r\right)^{n/3-\zeta_n}$ are not universal, as it
was also explicitly checked by direct comparison. Conversely, in
Fig.~4 it is shown that scaling exponents of even order moments are
the same for the two types of forcing. For the pdf's this means that,
although having different shapes, the curves are rescaling with $r$ in
the same way.

The picture emerging from these results is as follows\,: structure
function exponents $\zeta_n$ are universal, while constants, and thus
the pdf's of scalar increments, are not.  The difference between
isotropic and anisotropic situations is that the non-universal
constants $C_{2n+1}$ in odd-order structure functions vanish by
symmetry for the former case, while they generically do not for the
latter.  Structure functions present anomalous scaling and there is no
full restoration of isotropy while going toward small scales.  This
picture of universality is weaker than in Kolmogorov's 1941 theory,
but coincides with the one emerged for intermittency in the Kraichnan
passive scalar model \cite{CFKL95,GK95,SS95} (see also
Ref.~\cite{SS99}).  The velocity used here has finite correlation
times, scalar correlation functions do not obey closed equations, yet
the universality properties are the same. This points to a
broader validity of the mechanisms identified for the Kraichnan model
and it is likely that the same universality framework generically
applies to passive scalar turbulence.

Let us now discuss the consequences of cliffs for the
\narrowtext \begin{figure} \epsfxsize=9truecm \epsfbox{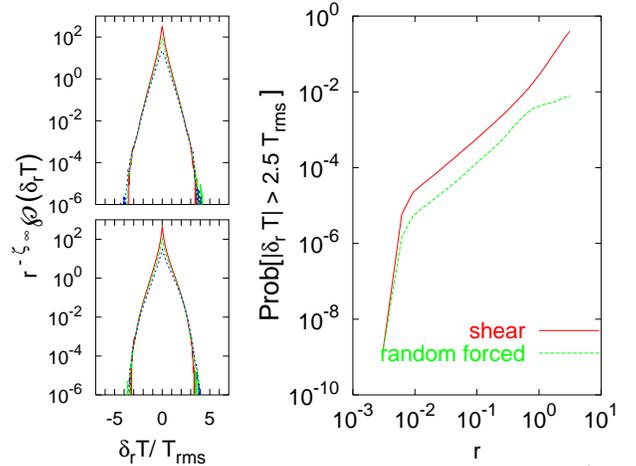}
\caption{On the left, the pdf's in Fig.~3 multiplied by
$r^{-\zeta_{\infty}}$. The upper curves are for the random forcing and
the lower ones for the shear case. On the right, cumulated
probabilities for scalar fluctuations to exceed $\lambda T_{rms}$ {\it
vs} $r$, with $\lambda=2.5$.}  \label{f6}
\end{figure}
\noindent intermittency at high orders. Their singularity strength
suggests that the scaling exponents might saturate at large orders,
i.e. $\zeta_n$ tends to a constant $\zeta_{\infty}$ for large enough
$n$. Physical self-consistency for the survival of steepening strong
fronts is demonstrated in Ref.~\cite{RHK97}, where saturation is shown
to imply that dissipation preferentially spares the cliffs with the
largest jumps.  High-order structure functions in our simulations are
shown in Fig.~5, together with the $\zeta_n$ {\it vs} $n$ curve,
compatible with saturation.  The same holds for ratios of two moments
{\it vs} $r$ or one moment {\it vs} the other. Note that, for any
finite-size field, there are orders where the moments start to be
spoiled and some strongest single structure having a dissipation width
will plausibly dominate the statistics, as in Burgers' equation
\cite{RHK99}. The convergence of the moments was inspected by the
usual test of checking that $(\delta_r T)^{14} {\cal P}$ decays before
the pdf ${\cal P}(\delta_r T)$ of the scalar 
\narrowtext \begin{figure} \epsfxsize=9truecm \epsfbox{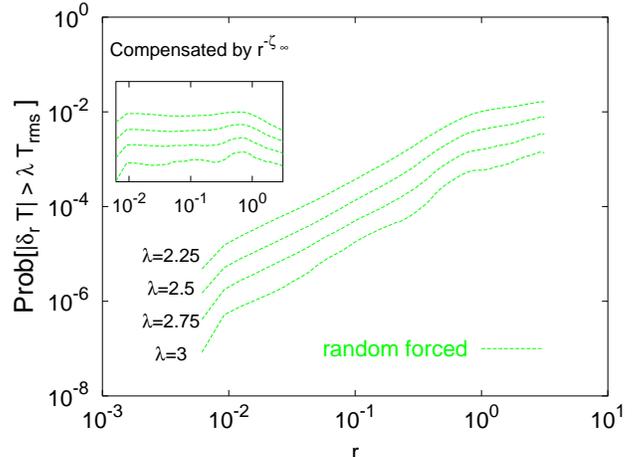}
\caption{As in Fig.~6, for different values of $\lambda$ in the randomly
forced case. The curves compensated by $r^{-\zeta_\infty}$, with
$\zeta_{\infty}=1.4$, are shown in the inset.}  \label{f7}
\end{figure}
\narrowtext \begin{figure} \epsfxsize=9truecm \epsfbox{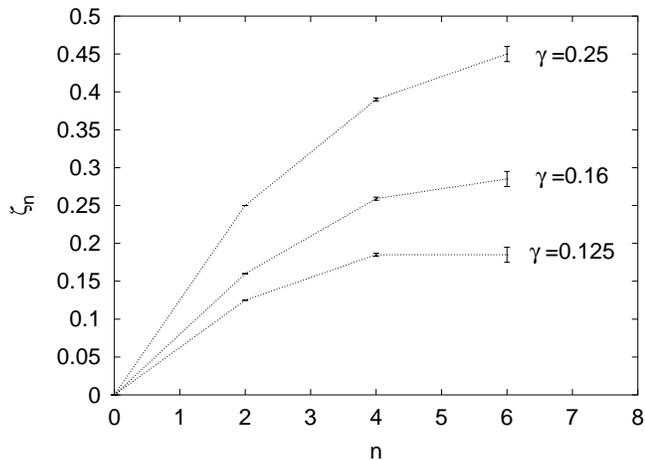}
\caption{Scaling exponents of the 2nd, 4th and 6th-order scalar
structure functions for three different roughness exponents $\gamma$
in the Kraichnan model.} \label{f8}
\end{figure} 
\noindent increments $\delta_r
T\equiv T(r)-T(0)$ becomes noisy. An alternative observable more
reliable than moments (as less sensitive to the extreme tails) is
looking, for fixed $\delta_r T$, at how ${\cal P}$ varies with the
separation $r$. Saturation is equivalent to the pdf taking the form
${\cal P}(\delta_r T)=r^{\zeta_\infty}{\cal Q}(\delta_r T/T_{rms})$
for $\delta_r T$ sufficiently larger than $T_{rms}=\langle
\left(T-\langle T\rangle\right)^2\rangle^{1/2}$. The collapse of the
curves $r^{-\zeta_\infty}{\cal P}(\delta_r T)$ in Fig.~6 is therefore
a signature of saturation and gives the unknown function ${\cal
Q}$. In Fig.~7, we plot the cumulated probabilities $\int_{\delta_r
T}^{\infty}{\cal P}$ {\it vs} $r$ for various $\delta_r T$ and the
parallelism of the curves is again the footprint of
saturation. Explicit evidence for the universality of $\zeta_{\infty}$
is provided in Fig.~6.

The physical origin of cliffs resides in the Lagrangian structure of
(\ref{scalar}), i.e. in the fact that particles are passively
transported by the velocity ${\bm v}$. In regions where velocity
gradients are sufficiently persistent in space and time, widely spaced
particles tend to approach and generate the observed abrupt variations
of the scalar field.  This suggests that, even though quantitative
aspects, such as the order of saturation or the value $\zeta_\infty$,
depend on the choice of ${\bm v}$, the saturation phenomenon itself
should occur for a wide class of random velocity fields. The Kraichnan
model \cite{RHK94} is unfavorable for saturation because of the short
velocity correlation time. Despite this, for large dimensionalities of
space, saturation analytically follows from an instanton solution
\cite{BL98}.  For the 3D case, saturation was phenomenologically
suggested in Ref.~\cite{VY} and inferred from an instantonic bound in
Ref.~\cite{MC}. Direct numerical evidence is provided by our 3D
numerical simulations whose results are presented in Fig.~8. Scaling
exponents have been measured using the Lagrangian method presented in
Ref.~\cite{FMV98} and $(2-\gamma)/2$ is the spatial H\"older exponent
of ${\bm v}$, as in Ref.~\cite{SS99}. The order of the moments needed
to observe saturation is expected to diverge for $\gamma\to 2$, while
for $\gamma\to 0$ the action of large-scale gradients should favor
close approaches between particles.  The order is thus expected to
reduce with $\gamma$ and for the smoothest velocity in Fig.~8
saturation is indeed occurring already at the 4th-order and thus
becomes observable. This confirms the physical picture of saturation
due to the cliffs formed in the scalar field and the genericity of the
phenomenon for scalar turbulence intermittency.

\medskip
{\bf Acknowledgements.} Helpful discussions with M.~Chertkov,
A.~Fairhall, U.~Frisch, B.~Galanti, R.H.~Kraichnan, V.~Lebedev,
A.~Noullez, J.F.~Pinton, I.~Procaccia and A.~Pumir are gratefully
acknowledged. We benefited from the hospitality of the 1999 ESF-TAO
Study Center. The INFM PRA Turbo (AC) and the ERB-FMBI-CT96-0974 (AL)
contracts are acknowledged. Simulations were performed at IDRIS
(no~991226) and at CINECA (INFM Parallel Computing Initiative).

\end{document}